\documentclass[aps,pre,twocolumn,showpacs,final,floatfix, longbibliography, superscriptaddress]{revtex4-2}
\usepackage{amsmath}
\usepackage{amsfonts}
\usepackage{amssymb}
\usepackage{physics}
\usepackage{color}
\usepackage{hyperref}
\usepackage{inputenc}
\usepackage{enumerate}
\usepackage{graphicx}
\usepackage{caption}
\usepackage{float}
\usepackage{amsmath}
\usepackage{amsfonts}
\usepackage{lipsum}
\usepackage{dcolumn}
\usepackage{hyperref}
\usepackage{subfigure}
\usepackage{epsfig}
\usepackage{epsf} 
\usepackage{epstopdf}

\usepackage{color}
\usepackage{enumitem}
\usepackage{xr}
\makeatletter

\newcommand*{\addFileDependency}[1]{
\typeout{(#1)}
\@addtofilelist{#1}
\IfFileExists{#1}{}{\typeout{No file #1.}}
}\makeatother

\usepackage{hyperref}
\hypersetup{colorlinks,
    citecolor=blue,
    filecolor=cyan,
    linkcolor=blue,
   urlcolor=magenta
 }

\usepackage{cleveref}

\newcommand {\apgt} {\ {\raise-.5ex\hbox{$\buildrel\rangle\over\sim$}}\ }
\newcommand {\aplt} {\ {\raise-.5ex\hbox{$\buildrel\langle\over\sim$}}\ }

\newcommand{\si}{\sigma}
\newcommand{\sib}{\bar{\sigma}}

\newcommand{\beq}{\begin{eqnarray}}
\newcommand{\eeq}{\end{eqnarray}}
\newcommand{\barray}{\begin{eqnarray}}
\newcommand{\earray}{\end{eqnarray}}
\newcommand{\nn}{\nonumber}

\newcommand{\up}{\uparrow}
\newcommand{\dn}{\downarrow}

\usepackage{hyperref}
\hypersetup{
    colorlinks,
    citecolor=red,
    filecolor=cyan,
    linkcolor=blue,
   urlcolor=magenta
 }

\usepackage{xparse}
\ExplSyntaxOn
\NewDocumentCommand{\mref}{m}{\quinn_mref:n {#1}}
\seq_new:N \l_quinn_mref_seq
\cs_new:Npn \quinn_mref:n #1
 {
  \seq_set_split:Nnn \l_quinn_mref_seq { , } { #1 }
  \seq_pop_right:NN \l_quinn_mref_seq \l_tmpa_tl
  ( 
  \seq_map_inline:Nn \l_quinn_mref_seq
    { \ref{##1},\nobreakspace } 
  \exp_args:NV \ref \l_tmpa_tl 
  ) 
 }
\ExplSyntaxOff

\newcommand{\disp}[1]{Eq.~\mref{#1}}

\begin{document}

\title{Partition function zeros of quantum many-body systems: perturbative results}


\author{Muhammad Sedik and  B Sriram Shastry\\
Physics Department, University of California, Santa Cruz, CA 95064}

\date{\today}

\begin{abstract}
A  systematic method to find the Yang-Lee partition function zeros of quantum many-body systems based on perturbation theory at finite temperatures was recently introduced \cite{Shastry2025Yang}.
This method identifies wave-vector and temperature-dependent complex virtual energies obtainable from the thermal electronic  Greens function. The collection of virtual energies over all $\vec{k}$ yield the Yang-Lee zeroes. 
 We apply this approach to the one-dimensional Hubbard model for different boundary conditions. We compare the results obtained by this method up to second-order in perturbation theory, with the results found by exact diagonalization. We also propose a quantity that could be used for experimental detection of these zeros of a Hubbard ring. An example of the detection method is presented using exact diagonalization of the 8-site Hubbard ring. 
\end{abstract}

\maketitle

\section{Introduction}

 A systematic method to find the partition function zeros of quantum many-body systems, using perturbation theory in the interaction strength, has been proposed recently  \cite{Shastry2025Yang}. Starting from the known zeros of the non-interacting system, this perturbative method tracks the evolution of these zeros systematically with interactions, thereby leading to an unexpected wave vector labeling of the zeros.

C. N. Yang and T. D. Lee's seminal work in this direction focused on classical many-body systems\cite{yang1952statistical} undergoing phase transitions. Amongst other general results,  they showed that the zeros of the partition function of an Ising ferromagnet lie on the unit circle in the complex fugacity plane \cite{lee1952statistical}.  This is the central result of the  Lee-Yang circle theorem. An important generalization to the quantum case of the anisotropic Heisenberg ferromagnet is due to Asano \cite{asano1970theorems},  Ruelle and Dyson\cite{ruelle1971extension}.  The method of contractions emerged from Asano's work, and is very effective for a certain class of problems, where it locates the region containing the zeroes of the partition function, { without} requiring an exact evaluation of the partition function.

For  classical systems, both analytical \cite{katsura1954phase,hemmer1964yang,heilmann1970monomers, heilmann1971location,lieb1972property,newman1974zeros,lieb1981general,biskup2000general,lebowitz2012location,li2023lee}, and numerical results are available\cite{suzuki1970statistical,katsura1971distribution,kim2004yang,deger2020lee,hwang2010yang,sedik2024yang}. In addition to experimental realizations of Yang-Lee zeros \cite{wei2012lee,peng2015experimental,brandner2017experimental,wei2017probing}, they have also been realized on quantum computers \cite{francis2021many}.  Recently, the interest in applying the Yang-Lee approach to quantum many-body systems has increased. Yang-Lee zeros of different quantum models have been studied through various methods, such as tensor-network \cite{kist2021lee,Liu2024Exact,Vecsei2025Lee}, renormalization group \cite{li2023yang}, quantum entanglement \cite{li2025yang}, among other \cite{brange2024lee,Gu2026Fidelity}.

The quantum many-body problems, such as the Hubbard model,  where the new method \cite{Shastry2025Yang} is applicable, are otherwise difficult to handle, since the method of contractions is not applicable. Numerical methods such as exact diagonalization are also difficult to apply in view of the rapid growth of the size of the Hilbert space with system size, as well as the high numerical precision required to obtain the zeros of the polynomials that arise  \cite{lin1993exact,jafari2008introduction,kingsley2013exact}.  Studies of the repulsive Hubbard model were made using Quantum Monte-Carlo (QMC) simulations \cite{barbour1992grand,abraham1996singularity}. However, the studies showed only the zeros closest to the real axis, seemingly due to QMC limitations, so the complete picture of the Yang-Lee zeros distribution of this model remains unknown. 

We summarize the method proposed by Shastry \cite{Shastry2025Yang} and provide a motivation for this work.
This method uses a theorem  \cite{Shastry2025Yang} involving the single electron self-energy found from the imaginary time Matsubara formalism. The theorem relates to  the inverse Green's function at the lowest (positive imaginary)  frequency $\omega_{n=0}=i \frac{\pi}{\beta}$ 
\begin{align}
    \Psi_{\vec{k} \sigma}(\mu) &\equiv G_\sigma^{(-1)}\left(\vec{k}, \left.i \frac{\pi}{\beta}\right|\mu\right)\notag\\
    &= \mu+ i \frac{\pi}{\beta}-\varepsilon_\sigma(\vec{k})- \Sigma_\sigma\left(\vec{k}, \left.i \frac{\pi}{\beta}\right|\mu\right)
\label{eqn:inverse-Greens-function}
\end{align}
in a domain ${\cal S}=\{\mu:-\frac{2 \pi}{\beta}< \Im  \mu \leq 0\}$. Here $\Sigma$ is the proper (Matsubara-Dyson) self-energy.
The  relevant theorem in Ref.~\cite{Shastry2025Yang} says that for a given $\{\vec{k} \sigma\}$, if one can find $\mu^*$ where  $\Psi_{\vec{k} \sigma}(\mu^*)=0$, then the (grand canonical)  partition function vanishes at $\mu^*$, i.e.  $Z(\mu^*)=0$.  From this theorem, it is argued that all the zeros of $Z$ can be found by studying the entire set of $2 N_S$ equations found from the two directions of spin $\sigma$ and the $N_s$ values of the crystal momentum $\vec{k}$. It is noteworthy that within this method, two zeros of the partition function can be associated with each wave vector $\vec{k}$. This feature  is illustrated below, and  allows us to define \cite{Shastry2025Yang} virtual energies  $\xi_{\vec{k} \sigma}$ that are closely related to the zeros $\mu^*$ from
\begin{equation}  
\xi_{\vec{k}\sigma}= \mu^* +i \frac{\pi}{\beta}- \frac{U}{2}
\label{eqn:xi-of-mu-and-U}
\end{equation}
 The partition function zeros now translate into a new band of (often complex) virtual energies, which reduce to the $T$-independent bare band structure energies $\varepsilon_\sigma(\vec{k})$ at $U=0$.

This formalism was illustrated in \cite{Shastry2025Yang} by applying it to the 1-dimensional Hubbard model with 6 sites using the first-order perturbation result for $\Sigma$, and comparing with the exact partition function zeros found from exact diagonalization. This study reveals interesting differences between attractive and repulsive signs of the interaction, and gives insights into the location of roots for the repulsive case. The present work applies the perturbative method to one higher order in perturbation theory, and brings out new features relating to the evolution of virtual energies and the role of certain sum-rules which these must satisfy on general grounds.  The exact diagonalization calculations are performed with up to  9  sites, which is almost the limit of sizes accessible at present. We also point out the interesting connection between time-dependent charge fluctuations in a ring of the Hubbard model atoms sitting on a charged 2-d layer, and its mapping to the zeros of the partition function with a complex chemical potential.

This paper is structured as follows. In Sec.~\ref{sec:2nd-order-results}, we introduce the model, present virtual energies obtained by ED, briefly introduce the inverse Green's function, and draw comparisons between ED, first-order perturbation, and second-order perturbation results for different boundary conditions. We propose a quantity that could be used for experimental detection of Yang-Lee zeros of the Hubbard model in Sec.~\ref{sec:charge-fluctuations}. We finally summarize our work in Sec.~\ref{sec:conclusion}.

\section{Virtual Energies From Self-energy up to $O(U^2)$}
\label{sec:2nd-order-results}
\subsection{Hubbard model and virtual energies}
\label{subsec:model-and-VE}
We study the 1-d Hubbard model on $N_s$ lattice points with the Hamiltonian
\begin{align}
    H=\sum_{k, \si=\pm1} \varepsilon_{k} C^\dagger_\si(k) C_\si(k) + U \sum_{i} n_{i \up} n_{i \dn},
    \label{eqn:Hubbard-H}
\end{align}
where $C_\sigma(k)$ is the fermionic annihilation operator, $    \varepsilon_{k}= -2 t \cos k$,
 where $k=\frac{2\pi \nu}{N_s}$ for  integer $\nu=0,1,\ldots N_s-1$ in  the first Brillouin zone $k\in [-\pi,\pi)$. 
The grand canonical partition function is then given by 
\begin{align}
    Z(z)=\tr\left(e^{-\beta H + \beta \mu \hat{N}}\right)=\sum_{j=0}^{2N_s}Z_j z^j,
    \label{eqn:GPF}
\end{align}
where $\hat{N}\equiv \sum_{k,\sigma}C^{\dag}_\sigma(k)C_\sigma(k)$ is the number operator, $Z_j$ is the canonical partition function for a system with $j$ particles, and the fugacity $z=e^{\beta \mu}$ with $\beta=1/k_B T$.  The Yang-Lee zeros $z_j$ are the roots of the grand canonical partition function $Z(z_j)=0$.

The polynomial in Eq.~(\ref{eqn:GPF}) can be written as
\begin{align}
    Z=c_0\prod_{j=1}^{2N_s}(z-z_j)=\prod_{k\sigma}\left[1+\frac{e^{\beta \mu}}{e^{\beta\left(\xi_{k\sigma}+U/2\right)}}\right],
    \label{eqn:GPF-virtual-energies}
\end{align}
where $\xi_{k\sigma}$ are  virtual energies  defined in \disp{eqn:xi-of-mu-and-U}. The Yang-Lee zeros are related to the virtual energies by the mapping  in Eq.~(\ref{eqn:xi-of-mu-and-U}),
\begin{align}
    \mu_j \rightarrow -\frac{i\pi}{\beta}+\frac{1}{2}U+\xi_{k\sigma} \Longleftrightarrow z_j \rightarrow -e^{\beta\left( \xi_{k\sigma}+U/2\right)}.
    \label{eqn:index-mapping}
\end{align}
 This mapping  connects the set of labels $j \leftrightarrow k\sigma$, and thereby the unlabeled Yang-Lee zeros $z_j$(=$e^{\beta \mu_j}$)  to the virtual energies $\xi_{k\sigma}$. 
For ensuring the uniqueness of the mapping, $\mu_j$ are restricted to the domain
\beq
\mathcal{S}=\{\mu_j:\frac{-2\pi}{\beta}<\Im \mu_j \leq 0, \mbox{  } \mbox{Re} \, \mu_j \in( - \infty,\infty)\},, \label{Domain-S} 
\eeq 
and correspondingly
the virtual energies will be restricted in the domain $\frac{-\pi}{\beta}<\Im \xi \leq \frac{\pi}{\beta}$ 

To directly determine the Yang-Lee zeros $z_j$ of the Hubbard model, we exactly diagonalize the Hamiltonian in Eq.~(\ref{eqn:Hubbard-H}) using the techniques explained in Ref.~\cite{sharma2015organization} up to $N_s=9$. Without loss of generality, we set $t=1$ and obtain all eigenvalues for different $U$ and $\beta$. We then compute the grand canonical partition function in Eq.~(\ref{eqn:GPF}) and find its zeros $z_j$ using the {\em Mathematica} polynomial-root solver. We then find the set of virtual energies $\{\xi_{k\sigma}\}$ by computing $-U/2+\ln\left(-z_j\right)/\beta$ from the set of  $z_j$. It should be clear that the assignments of the wavevector and spin labels $\vec{k},\sigma$ are not possible within the exact diagonalization method. 

\begin{figure}[H]
    \centering
    \includegraphics[width=\columnwidth]{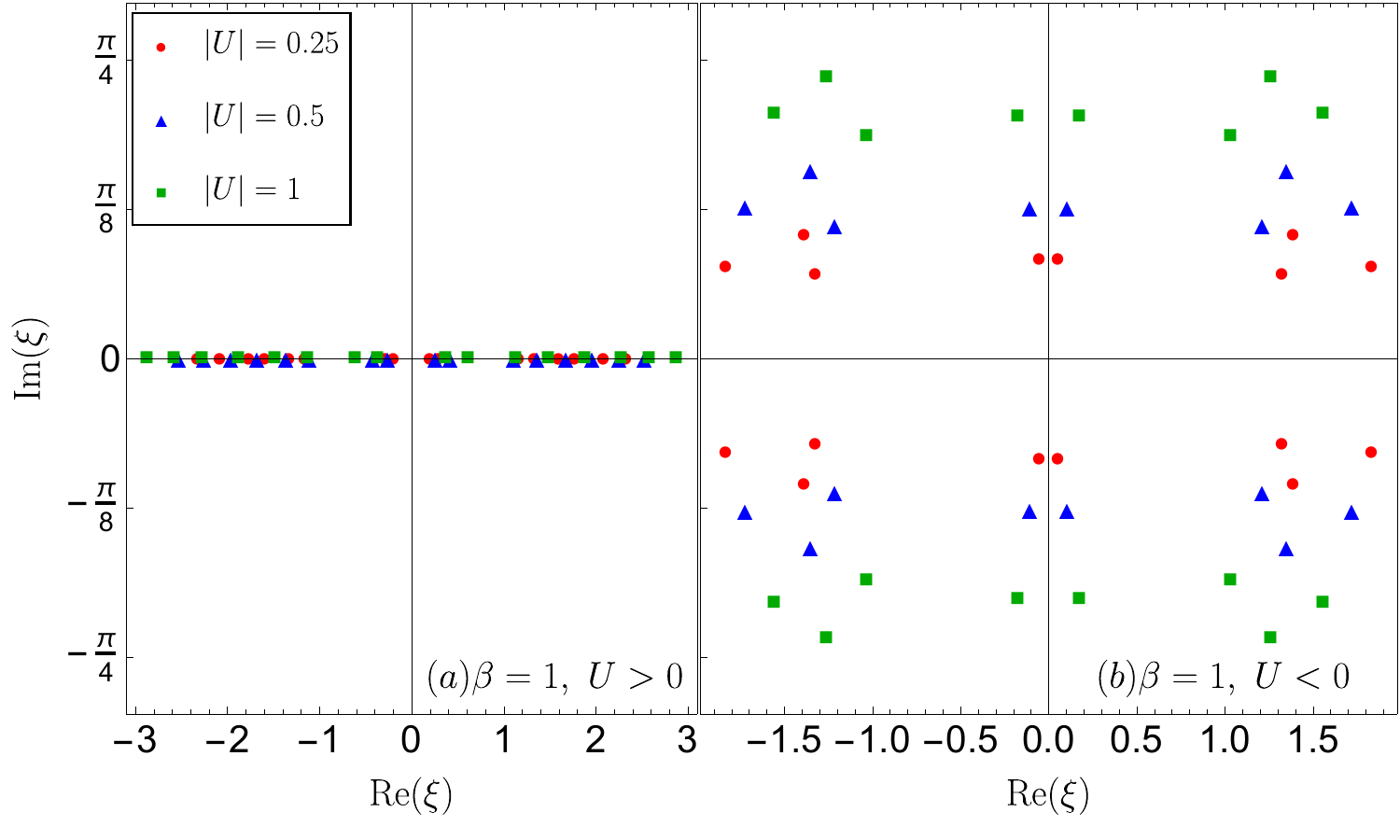}
    \caption{\footnotesize The virtual energies in the complex $\xi$-plane for $N_s=8$ lattice with periodic boundary conditions at $\beta=1$ and $t=1$, and different values of on-site coupling $U$ for: (a)$~U>0$ and (b)$~U<0$. $|U|=0.25$ is shown in red dots, $|U|=0.5$ in blue triangles, and $|U|=1$ in green squares. The virtual energies are symmetric around $\Re(\xi)=0$, and the distance between the virtual energies with the smallest $|Re(\xi)|$ is increasing with $|U|$, showing an increasing gap around this axis. As $U\rightarrow 0$, the zeros approach the values $\varepsilon_{k\sigma}$. For $U>0$, the roots lie on the real line $\Im  \left(\xi\right)=0$.}
    \label{fig:ED_roots}
\end{figure}

Fig.~\ref{fig:ED_roots} shows virtual energies for different on-site interactions $U$ found by exact diagonalization. For the case of a repulsive interaction $U>0$ Fig.~\ref{fig:ED_roots}(a), all the virtual energies lie on the real axis in the $\xi$-plane -- an observation that persists for all values of $U>0$ that we studied and all system sizes up to $N_s=9$. On the other hand, the virtual energies for an attractive interaction $U<0$ Fig.~\ref{fig:ED_roots}(b) are all complex-valued, and they move away from the real axis and closer to the lines $\Im \xi = \pm \frac{\pi}{\beta}$ upon increasing $|U|$. For all $U$ and even $N_s$, the virtual energies are symmetric around the imaginary axis $\Re(\xi)=0$.

In the next subsection, we review the perturbative approach to find Yang-Lee zeros introduced in  \cite{Shastry2025Yang} and extend the method to find the zeros up to $O(U^2)$.  

\subsection{The perturbative approach}
\label{subsec:pertrubation}
We start by noting the usual formula for the inverse Green's function in terms of the Matsubara frequency $\omega_n= (2 n+1)\pi/\beta$
\begin{align}
    G^{-1}_{\si}(k, i \omega_n)= i \omega_n+ \mu-\varepsilon_{k \si}- \Sigma_\si(k, i \omega_n),
    \label{eqn:inverse-Greens}
\end{align}
and denote the fermi function for an electron as  $f_{k \si}= \{e^{\beta \left(\varepsilon_{k \si}- \mu\right)}+1\}^{-1}$
and the hole function $\bar{f}_{k\si}= 1- f_{k \si}$.  We will retain the spin dependence of  $\varepsilon_{k\sigma}$  for a few steps, in order to clarify the statement of the $N_s$ equations, which will admit $2 N_s$ solutions for  $\xi's$ as shown below.  The self-energy can be written out in perturbation theory as follows.
 \begin{figure}[h]
\centering
\includegraphics[width=.7\columnwidth]{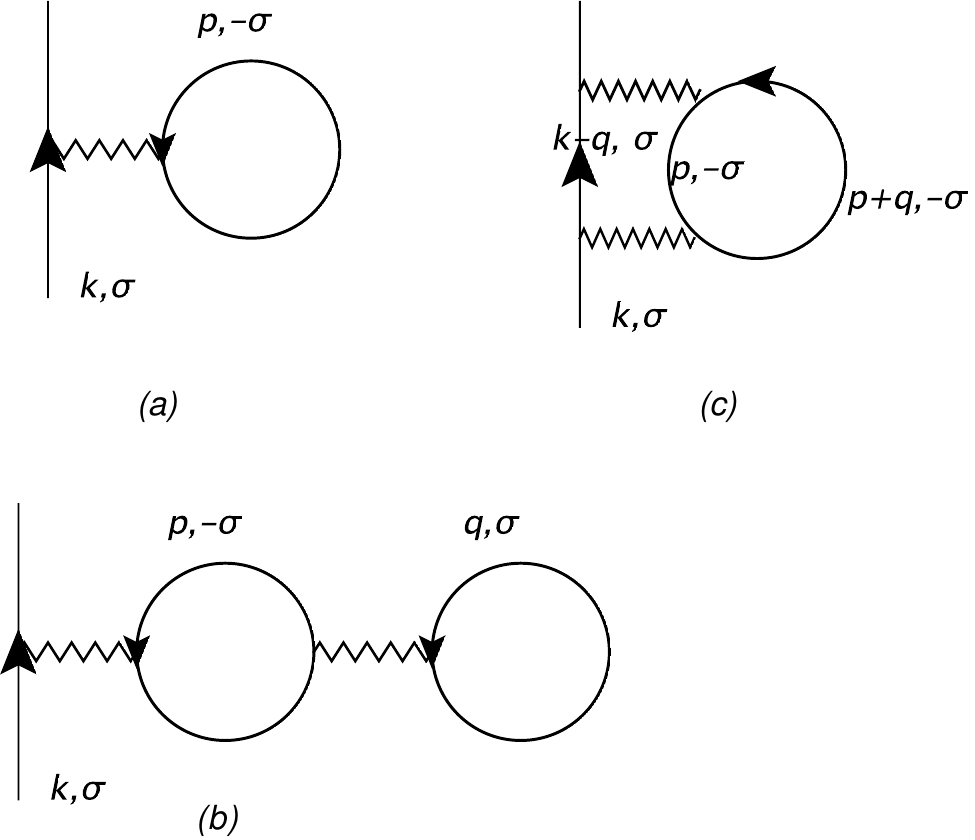}
 \caption{\footnotesize Feynman diagrams associated with first order in Eq.~(\ref{eqn:self-energy-1st}) and second in Eq.~(\ref{eqn:self-energy-2nd}) of perturbation theory of the self energy. As usual, solid lines represent $G$, and the wavy line represents the Coulomb-Hubbard interaction. Note that the spin label $-\sigma$ is denoted for brevity as $\bar{\sigma}$ in \disp{eqn:self-energy-1st} onwards.}
 \label{fig:Feynman-Diagrams}
 \end{figure}
Let us also note the expressions for the self-energy:
\begin{itemize}
\item First order [Fig.~\ref{fig:Feynman-Diagrams} (a)]
\begin{align}
    \Sigma^{(1)}_\si(k, i \omega_n) = \frac{U}{N_s} \sum_p f_{p \sib}
    \label{eqn:self-energy-1st}.
\end{align}

\item Second order [Fig.~\ref{fig:Feynman-Diagrams}(b) and (c)]
\begin{align}
    \Sigma^{(2)}_\si(k, i \omega_n) = \frac{U^2}{N^2_s} \sum_{p,q} \frac{\cal A}{\cal B}-\beta \frac{U^2}{N^2_s} \sum_{p,q}f_{p \sib}(1-f_{p \sib}) f_{q \si},
    \label{eqn:self-energy-2nd}
\end{align}
where
\begin{align}
{\cal A} &=f_{p+q \sib}  \bar{f}_{p \sib} f_{k-q \si}+ \bar{f}_{p+q \sib}  {f}_{p \sib} \bar{f}_{k-q \si}\label{eqn:cal-A},\\
{\cal B}&= i \omega_n +\mu+\varepsilon_{p+q \sib}-\varepsilon_{p \sib}-\varepsilon_{k-q \si}\label{eqn:cal-B}.
\end{align}
\end{itemize}
Since we are interested in using the inverse Green's function to locate Yang-Lee zeros, we set $n=0 \implies \omega_0=\pi/\beta$ and write $\mu=\xi-i\frac{\pi}{\beta}+\frac{1}{2}U$ to define
\begin{align}
f_{p \si}&=\frac{1}{1+e^{\beta(\varepsilon_{p \si}-\mu)}}\to&f_{p \si}(\xi)=\frac{1}{1-e^{(\beta\varepsilon_{p \si}-\xi-\frac{1}{2}U)}}. 
\label{eqn:fermi-function}
\end{align}
Hence, the zeros of inverse Green's function in Eq.~(\ref{eqn:inverse-Greens}) in the complex $\xi$ plane can be found up to second order in self-energy at a given $k$ and $\sigma$ from
\begin{align}
\xi_{k\sigma}=&\varepsilon_{k \sigma}+ \frac{U}{N_s}\sum_p f_{p \bar{\sigma}}(\xi_{k\sigma})\notag\\
&+ \frac{U^2}{N_s^2}\sum_{p q} \left\{\frac{{\cal A}}{{\cal B}}- \beta f_{p\bar{\sigma}}(\xi_{k\sigma})\left[1-f_{p \bar{\sigma}}(\xi_{k\sigma})\right] f_{q \sigma}(\xi_{k\sigma}) \right\}. \quad \quad
\label{eqn:YL-zeros-self-consist}
\end{align}
 \begin{figure}[H]
    \centering
    \includegraphics[width=\columnwidth]{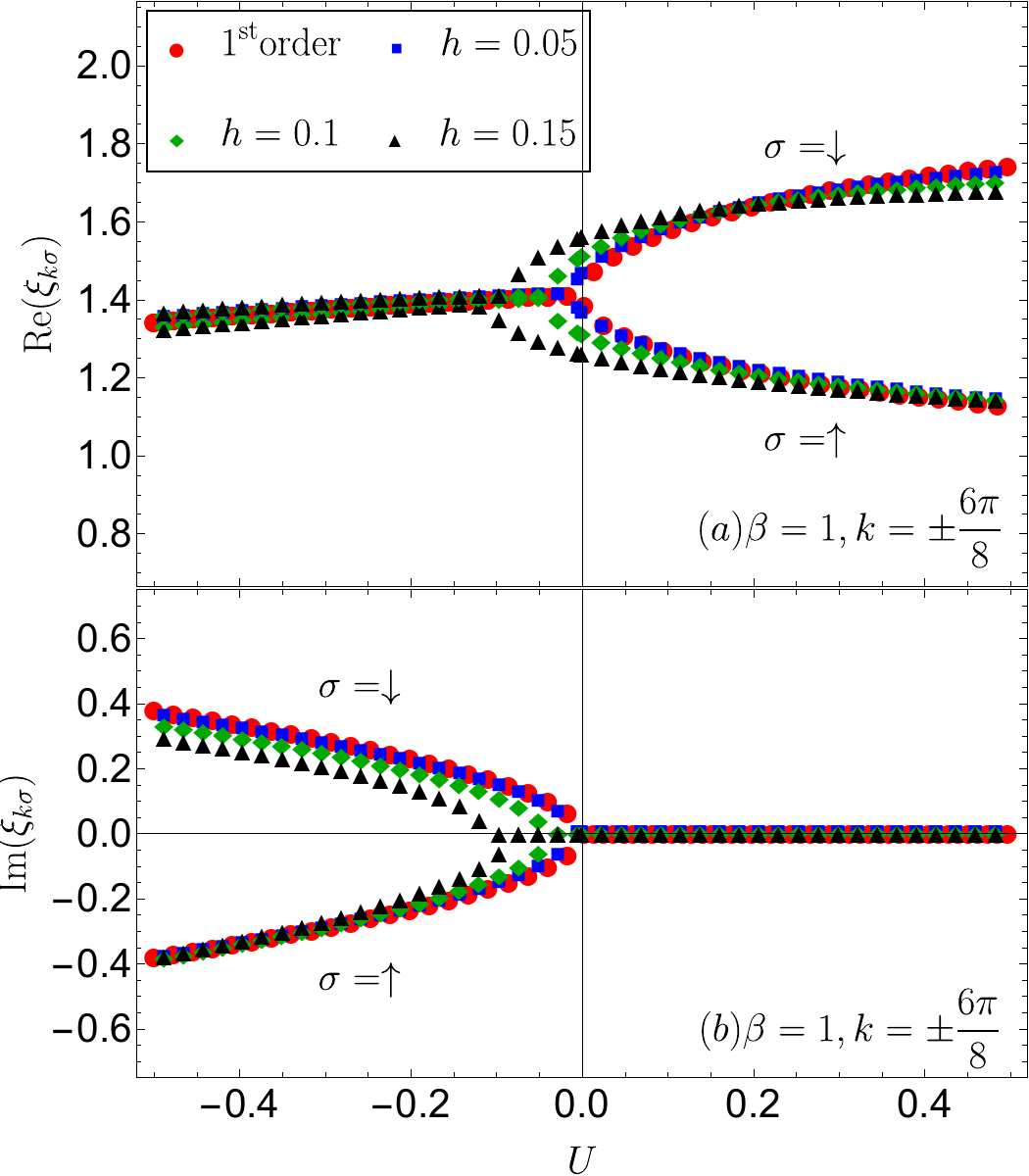}
    \caption{\footnotesize The virtual energies as a function of the interaction $U$ labeled by $k=\pm\frac{6\pi}{k}$ for $N_s=8$ lattice with periodic boundary conditions at $\beta=1$. The plot shows different values of the Zeeman field $h$, with the upper(lower) branch corresponding to $\sigma = \uparrow(\downarrow)$. It is observed that as $h \rightarrow 0$, both $\Re(\xi_{k\sigma})$ in (a) and $\Im(\xi_{k\sigma})$ in (b) approach the results found by the presented method (shown in red dots).}
    \label{fig:Zeeman_field}
\end{figure}
The sums over $p,q$ are with fixed spins as given without implying a spin summation.  This is a self-consistent equation for $\xi$, since it appears on both sides of the equation. It can be solved for a fixed $k$  numerically,
after ignoring the $\sigma$ dependence of  $\varepsilon_{k \sigma} $.
 Two solutions for $\xi$, which are  closest  to $\varepsilon_k$,
 are then  assigned $\sigma$ labels, as discussed below and in Ref.~\cite{Shastry2025Yang}. By solving Eq.~(\ref{eqn:YL-zeros-self-consist}) for all values of $k$, one should obtain the $2 N_s$ virtual energies $\xi$'s and hence the Yang-Lee zeros $z_j$. Another possible but tedious route, which is not taken here, would be to assign a $\ sigma$-dependence to $\varepsilon_{k \sigma}$ using a small Zeeman coupling and then take the limit of a vanishing field. We have checked cases where this leads to the same answers as the present method. We present results in Fig.~\ref{fig:Zeeman_field} for a test case of 8 sites with a finite Zeeman field $h$ and a variable $U$, where it is seen that the pair zero field results coincides with the results from the limit of finite field results found by the tedious route.
 
For the equation to be accurate up to $O(U)$ or $O(U^2)$, and given the explicit factors of $U$ in front,  one must expand the Fermi function in Eq.~(\ref{eqn:fermi-function}) around $U=0$ up to the desired order. Upon doing so, the self-consistent equation up to first order in $U$, given explicitly in $\xi_{k\sigma}$, is
\begin{align}
     \xi_{k\sigma}
        =\varepsilon_{k}&+ \frac{U}{2N_s}\sum_{p}\coth{\left[\frac{\beta}{2}\left(\xi_{k\sigma}-\varepsilon_k \right)\right]}+O(U^2),
        \label{eqn:theorem-1st-order-xi}
\end{align}
 and to the second order  in $U$ by
\begin{widetext}
    \begin{align}
         \xi_{k\sigma}
        =\varepsilon_{k}&+ \frac{U}{2N_s}\sum_{p}\coth{\left[\frac{\beta}{2}\left(\xi_{k\sigma}-\varepsilon_p \right)\right]}+\frac{\beta U^2}{8N_s^2}\sum_{p,q}\coth{\left[\frac{\beta}{2}\left(\xi_{k\sigma}-\varepsilon_p \right)\right]}\sinh^{-2}{\left[\frac{\beta}{2}\left(\xi_{k\sigma}-\varepsilon_q \right)\right]}\notag\\
        &+\frac{U^2}{N^2_s}\sum_{p,q} \frac{e^{\beta\left(\xi_{k\sigma}-\varepsilon_{p}\right)}-e^{\beta\left(2\xi_{k\sigma}-\varepsilon_{p+q}-\varepsilon_{k-q}\right)}}{\left[e^{\beta\left(\xi_{k\sigma}-\varepsilon_{p}\right)}-1\right]\left[e^{\beta\left(\xi_{k\sigma}-\varepsilon_{p+q}\right)}-1\right]\left[e^{\beta\left(\xi_{k\sigma}-\varepsilon_{k-q}\right)}-1\right]\left[\xi_{k\sigma}+\varepsilon_{p}-\varepsilon_{p+q}-\varepsilon_{k-q }\right]}+O(U^3).
        \label{eqn:theorem-2nd-order-xi}
    \end{align}
\end{widetext}
Note that we dropped the $\epsilon$ dependence on spin $\sigma$ here.  As discussed in \cite{footnote},
for every value of $k$, it is possible to find two solutions, which are identified as belonging to the two spin labels $\sigma=\pm1$

We present the solutions of Eqs.~(\ref{eqn:theorem-1st-order-xi},\ref{eqn:theorem-2nd-order-xi}) in the following subsection. 

\subsection{Virtual energies for periodic BCs}
\label{subsec:VE-periodic}
\begin{figure}[b]
    \centering
    \includegraphics[width=\columnwidth]{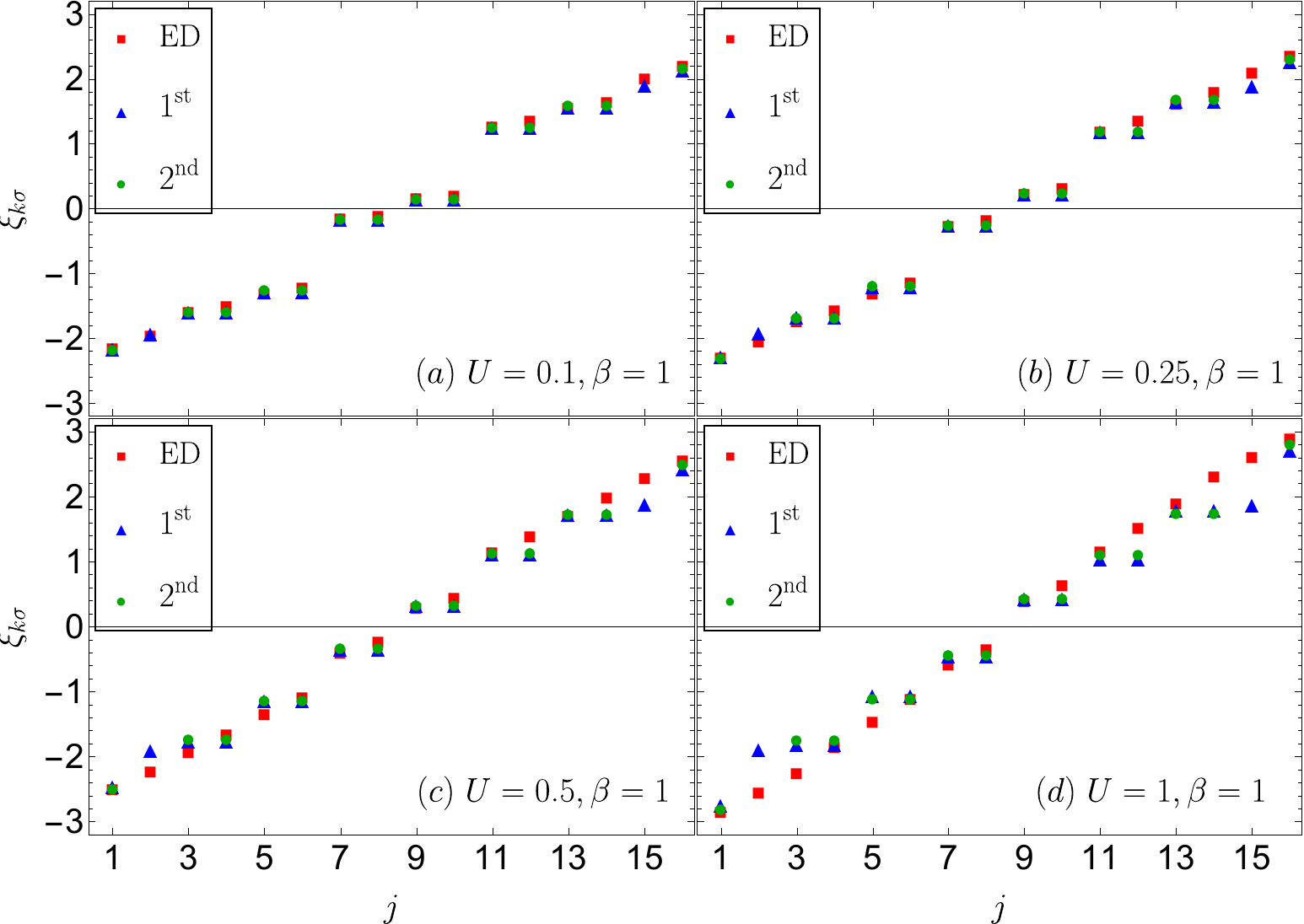}
    \caption{\footnotesize Comparison between the virtual energies $\xi_{k\sigma}$ of the 8-site Hubbard model with periodic BC's found by ED(red squares), Eq.(\ref{eqn:theorem-1st-order-xi})(blue triangles), and Eq.(\ref{eqn:theorem-2nd-order-xi})(green dots) at $\beta=1$ and different values of repulsive interaction $U>0$ marked on each plot. Since the ED virtual energies are all real for all $U\geq 0$ and $\beta$ we studied, we plot them on the vertical axis for easier comparison. The horizontal axis represents the mapping discussed below Eq.(\ref{eqn:index-mapping}), where the unlabeled ED virtual energies with index $j$ could be labeled with the quantum numbers $k\sigma$. For instance, the index $j=3$ corresponds to the label $k=\frac{\pi}{4}$ and $\sigma=\uparrow$, while the index $j=5$ corresponds to the label $k=\frac{\pi}{4}$ and $\sigma=\downarrow$ according to our convention. The comparison shows that the second-order $\xi$'s are closer to ED than the first-order $\xi$'s for most labels. However, both sets of solutions worsen with increasing $U$. The missing virtual energies at $k\sigma=-\pi\uparrow(j=2)$ and $k\sigma=0\downarrow(j=15)$ are due to the virtual energy real solution splitting into two complex-conjugate solutions around $U\approx0.022$.}  
    \label{fig:perturb_roots_periodic_repulsive}
\end{figure}
For every $k$ and $U$, we choose an iterative approach to solve Eqs.~(\ref{eqn:theorem-1st-order-xi},\ref{eqn:theorem-2nd-order-xi}). We track the evolution of virtual energies $\xi_{k\sigma}$ as the interaction strength \( U \) increases from zero up to a target value. For the first increment away from $U=0$, we pick the two solutions found by exploring starting points in a small (complex) neighborhood of  $\varepsilon_k$, and iterate from these until self-consistency is achieved.
Virtual energies found are either a real pair or a complex conjugate pair. In either case, we then iterate from this solution and find the next solution at a slightly larger U. This process ensures that the computed energy roots remain on smooth trajectories.
In a few cases, it transpires that one of the solutions is real and the other is complex. In this somewhat rare set of cases, we keep the real solution, the complex one, and its complex conjugate, leading to a number of solutions higher than $2N_s$, and discuss these cases separately below. 
\begin{figure}[h!]
    \centering
     \includegraphics[width=\columnwidth]{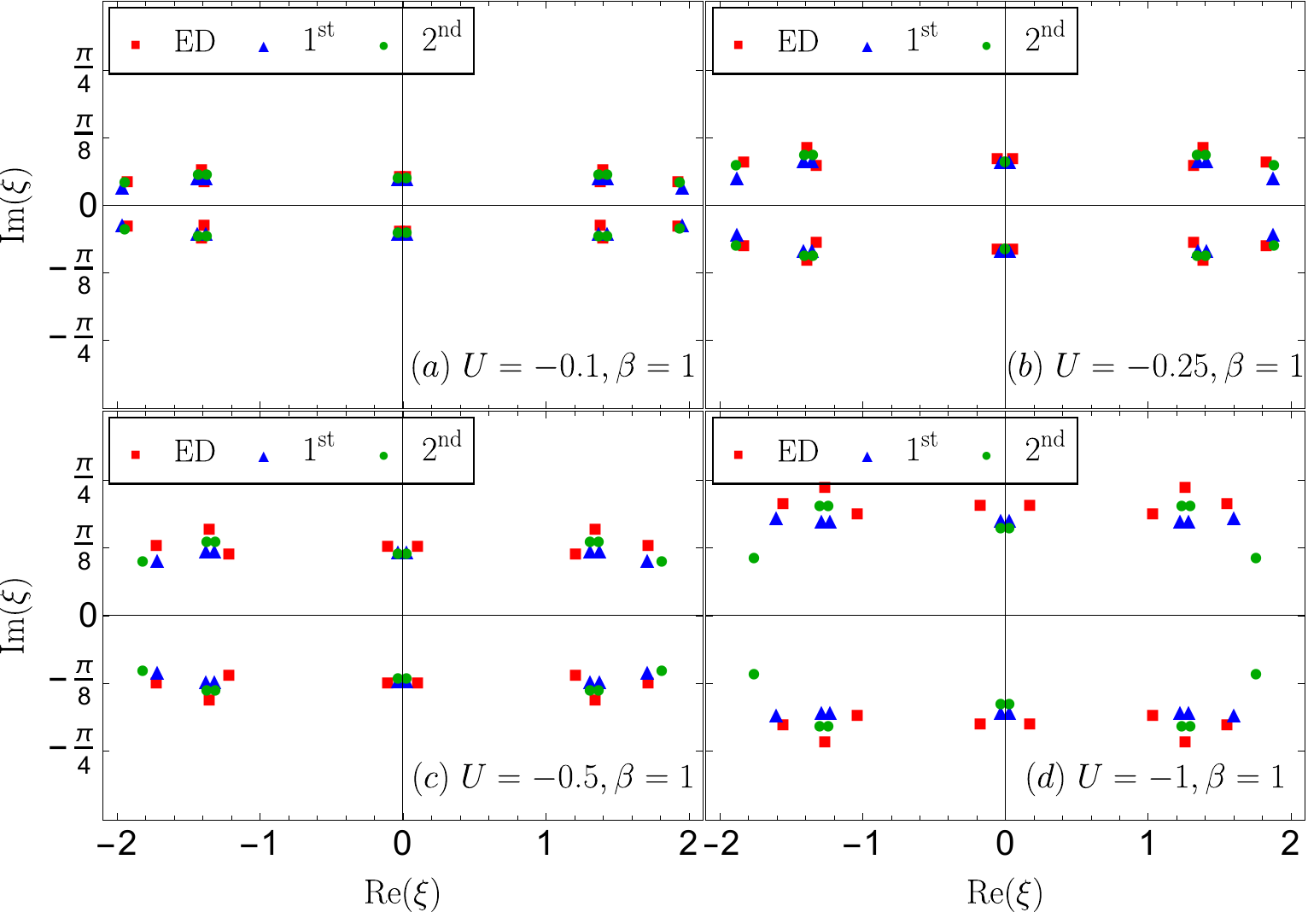}
    \caption{\footnotesize Comparison between the virtual energies $\xi_{k\sigma}$ of the 8-site Hubbard model with periodic BC's found by ED(red squares), Eq.(\ref{eqn:theorem-1st-order-xi})(blue triangles), and Eq.(\ref{eqn:theorem-2nd-order-xi})(green dots) at $\beta=1$ and different values of attractive interactions $U<0$ marked on each plot in the complex $\xi$-plane. For the degenerate values of momentum, namely $k=\pm\frac{\pi}{4},\pm\frac{\pi}{2},\pm\frac{3\pi}{4}$, the $\xi$'s are plotted such that a slight symmetric displacement is added to the real part. The comparison shows that the second-order $\xi$'s are closer to ED than the first-order $\xi$'s for most values of momentum, with the exception of $k=-\pi$ and $k=0$, which are at the extremities of the non-interacting energy band. In general, both sets of solutions worsen with increasing $|U|$.}
    \label{fig:perturb_roots_periodic_attractive}
\end{figure}

Figs. \ref{fig:perturb_roots_periodic_repulsive} and \ref{fig:perturb_roots_periodic_attractive} show a comparison between the virtual energies $\xi_{k\sigma}$ of the 8-site Hubbard model with periodic BC's found by ED, Eq.(\ref{eqn:theorem-1st-order-xi}), and Eq.(\ref{eqn:theorem-2nd-order-xi}) at $\beta=1$ and different values of repulsive interaction $U>0$ and attractive interaction $U<0$, respectively. For most data points, the second-order virtual energies are closer to ED than the first-order virtual energies, and both worsen with increasing $|U|$. However, the second-order $\xi$'s corresponding to the momentum labels $k=-\pi,0$ (edges of the energy band) are problematic in both signs of interaction. For $U>0$, the virtual energy real solution splits into two complex-conjugate solutions around $U\approx0.022$, whereas for $U<0$, the virtual energies move further away from ED as $|U|$ increases compared to the behavior of the first-order $\xi$'s.    

There are two remarks to note about the second-order $\xi$'s shown in Figs. \ref{fig:perturb_roots_periodic_repulsive} and \ref{fig:perturb_roots_periodic_attractive}: (1) the seeming discontinuity in the $\xi$'s corresponding to the momentum labels $k=-\pi,0$ for $U>0$, and (2) the degeneracy of $\xi$'s for $k=\pm\frac{\pi}{4},\pm\frac{\pi}{2},\pm\frac{3\pi}{4}$ that persists for both first- and second-order results. We discuss the former below while addressing the second in the next subsection. 

\begin{figure}[h!]
    \centering
    \includegraphics[width=\columnwidth]{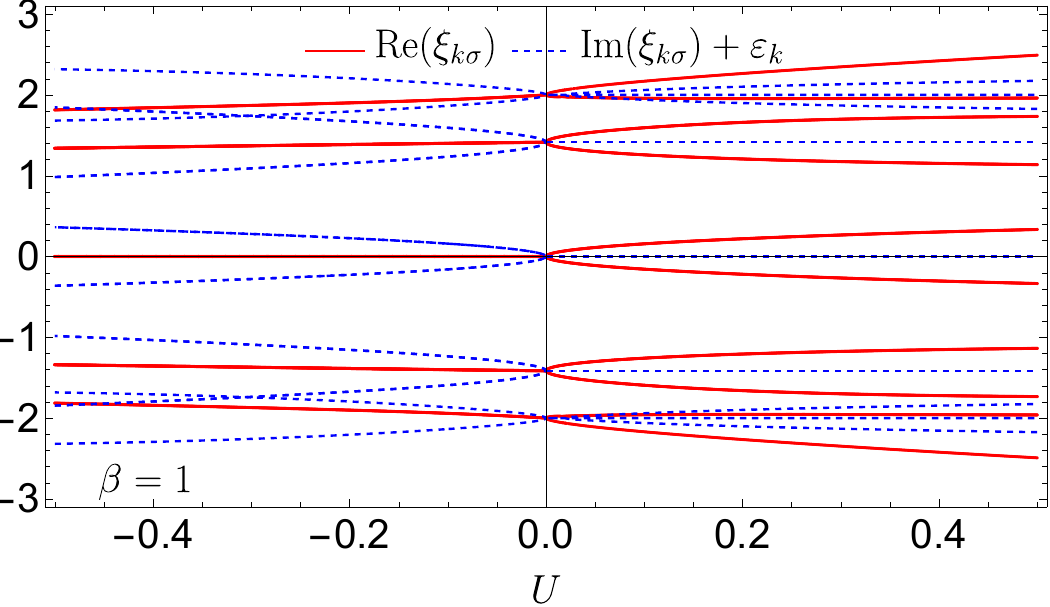}
    \caption{\footnotesize Plot of $\Re(\xi_{k\sigma})$ (solid red) and $\Im(\xi_{k\sigma})$ (dashed blue) as a function of $U$. The plot shows, from top to bottom, $k=-\pi,\pm\frac{3\pi}{4},\pm \frac{\pi}{2},\pm \frac{\pi}{4},0$, respectively. The imaginary parts are shifted by $\varepsilon_{k}$ for easier visualization at $U=0$. For values of $k$ where $\varepsilon_k$ is degenerate (bulk of the band), the solutions show as complex-conjugate pairs for $U<0$ approaching $\varepsilon_k$ at $U=0$ and turning into a pair of two real solutions. For the non-degenerate energy levels $\varepsilon_{-\pi}=2$ and $\varepsilon_{0}=-2$, one of the two real solutions splits into two complex-conjugate solutions indicated by the three dashed blue lines for $U\gtrapprox 0.022$.}
    \label{fig:xi-of-U-evolution_periodic}
\end{figure}

We study the continuity of the second-order $\xi$'s by plotting $\Re(\xi_{k\sigma})$ and $\Im(\xi_{k\sigma})$ as a function of $U$ as shown in Fig. \ref{fig:xi-of-U-evolution_periodic}. The continuity in both real and imaginary parts for $\xi$'s is evident in the plot. Each branch corresponds to a given value of $k$, splits into either two real solutions $U>0$ or two complex-conjugate solutions $U<0$ signaling the $\sigma=\pm 1$ labels. This gives a total number of $2N_s$ virtual energies after accounting for the degeneracy in $\varepsilon_k$. However, $\xi_{-\pi\uparrow}$ and $\xi_{0\downarrow}=2$ split into two complex-conjugate pairs at $U\approx 0.022$, increasing the number of solutions to $2N_s+2$ for $U\gtrapprox 0.022$ because $k=-\pi,0$ now result in three virtual energies: one real and a complex conjugate pair.

The reason for keeping the three solutions is two fold: (1) one would like to keep the continuity of the real solution corresponding to $\xi_{-\pi\downarrow}(\xi_{0\uparrow})$ and (2) the real solution $\xi_{-\pi\uparrow}(\xi_{0\downarrow})$ turns into a complex-conjugate pair, but their real part forms a continuation of the original single real solution. The latter occurs because the right-hand side of Eq.~(\ref{eqn:theorem-2nd-order-xi}) has a second-order pole around $\xi=\varepsilon_k$, rather than a third-order one as in the case with the other values of $k\sigma$. The equation with a second-order pole has three real solutions, two of which are on the same side of the pole, but only one of them is $\xi_{-\pi\uparrow}(\xi_{0\downarrow})$. As $U$ increases, these two solutions merge into a single real solution, then into a complex-conjugate pair.

\begin{figure}[h!]
    \centering \includegraphics[width=\columnwidth]{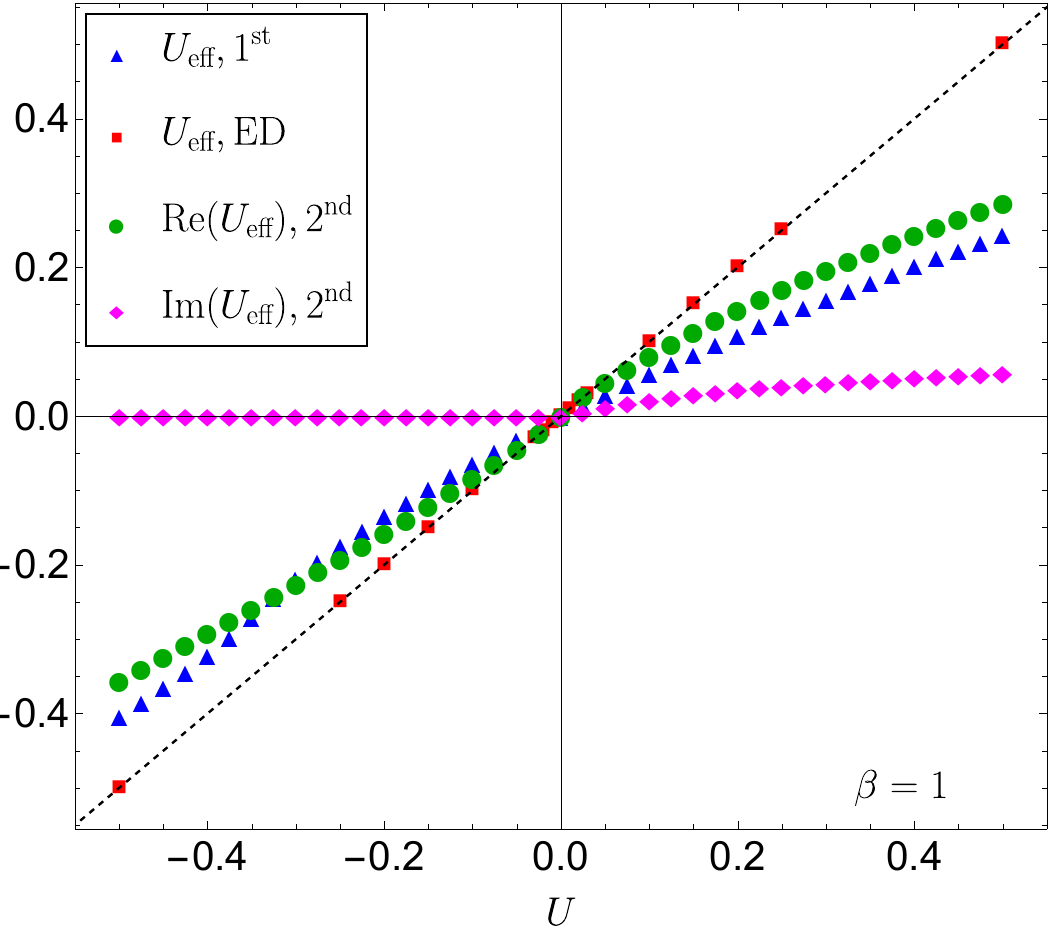}
    \caption{\footnotesize Plot of $U_\text{eff}(U,\beta,\left\{\xi\right\})$ for the different sets of $\xi$'s: ED (red squares), first order (blue triangles), and second order (green dots and purple diamonds). The black dotted line $U_\text{eff}=U$ is a guide for the eye. We see that the ED results lie on the dotted line as expected, while the first and second order approximations go further from the line with increasing $|U|$. For small enough $|U|$, the second order results are closer to the dotted line than the first order results, indicating a better quality of approximation, i.e., the $\xi$'s found perturbatively are globally closer to the exact virtual energies with increasing order $O(U)$.}
    \label{fig:U-eff-periodic}
\end{figure}

 We next quantify the quality of our first-order and second-order approximations using sum-rules derived in $\cite{Shastry2025Yang}$. By comparing the sides of Eq.~(\ref{eqn:GPF-virtual-energies}), one can find different sum-rule on the virtual energies. One of them is found from the coefficient of the single-particle canonical partition function $Z_1$, namely 
\begin{align}
    e^{-\frac{1}{2}\beta U}\sum_{k\sigma}e^{-\beta \xi_{k\sigma}}=\sum_{k\sigma}e^{-\beta \varepsilon_{k\sigma}}.
    \label{eqn:sum-rule-original}
\end{align}
Defining, 
\begin{align}
    U_\text{eff}(U,\beta,\left\{\xi\right\})=\frac{2}{\beta}\ln\left(\frac{\sum_{k\sigma}e^{-\beta \xi_{k\sigma}}}{\sum_{k\sigma}e^{-\beta \varepsilon_{k\sigma}}}\right),
    \label{eqn:U_eff}
\end{align}
the sum-rule \disp{eqn:sum-rule-original} can be written as 
\begin{equation} U_\text{eff}(U,\beta,\left\{\xi\right\})=U. 
\label{Ueff}
 \end{equation} 
Hence, for a set of numerically found $\xi$'s, one can measure the quality of this approximation to the exact $\xi$'s by checking the difference between $U_\text{eff}(U,\beta,\left\{\xi\right\})$ and $U$. 

Using the sum-rule to check the ED virtual energies and first order is straightforward. However, for the second-order solutions, the energies $\xi_{-\pi\uparrow}$ and $\xi_{0\downarrow}$ undergo complex-conjugate splitting. To maintain a single value for every $\xi_{k\sigma}$, we use the convention introduced earlier \cite{footnote}. This convention introduces a small imaginary part to $U_\text{eff}$ as the virtual energies are no longer appearing in complex conjugate pairs. 

A plot of $U_\text{eff}(U,\beta,\left\{\xi\right\})$ for the different sets of virtual energies (ED, first order, and second order) is shown in Fig. \ref{fig:U-eff-periodic}. The figure shows that even though $U_\text{eff}$ is complex for the second-order solutions, the real part is still closer to the exact value of $U$ (dotted black line) for small enough $|U|$. This suggests that the $\xi$'s found perturbatively are globally closer to the exact virtual energies with increasing order $O(U)$.  

In the next subsection, we discuss the degeneracy of $\xi$'s for degenerate values of $\varepsilon_k$ that persists for both first- and
second-order results. 

\subsection{Virtual energies for Twisted BCs}
\label{subsec:VE-twisted}
We study the effect of changing the degeneracy of the band energies on the first and second order Eqs.~(\ref{eqn:theorem-1st-order-xi}) and (\ref{eqn:theorem-2nd-order-xi}). This is achieved by considering different sets of boundary conditions.  If we impose a twist in the boundary condition of the form $C_{N+1,\sigma}=e^{i\phi}C_{1,\sigma}$ on the real-space hopping version of the hopping Hamiltonian, different values of $\phi$ correspond to different  BC's.  The band  energy dispersion with $\phi$ is 
\begin{align}
\varepsilon_k=-2t\cos{\left(k+\frac{\phi}{2\pi}\right)},
\label{eqn:tigh-binding-energies-twisted}
\end{align}
where $k=\frac{2\pi \nu}{N_s}$ for appropriate $N_s$ values of integer $\nu$ so that  $k$  is restricted to the first Brillouin. In the results shown above, where   $N_s$ is even, and periodic BC's, i.e., $\phi=0$ is used, there are two non-degenerate energy levels ($\varepsilon_0$ and $\varepsilon_{-\pi}$), while all other levels are doubly degenerate.  We now focus on  $\phi=\pi/2$, where all 
band energies are non-degenerate.

\begin{figure}[h!]
\centering
    \includegraphics[width=\columnwidth]{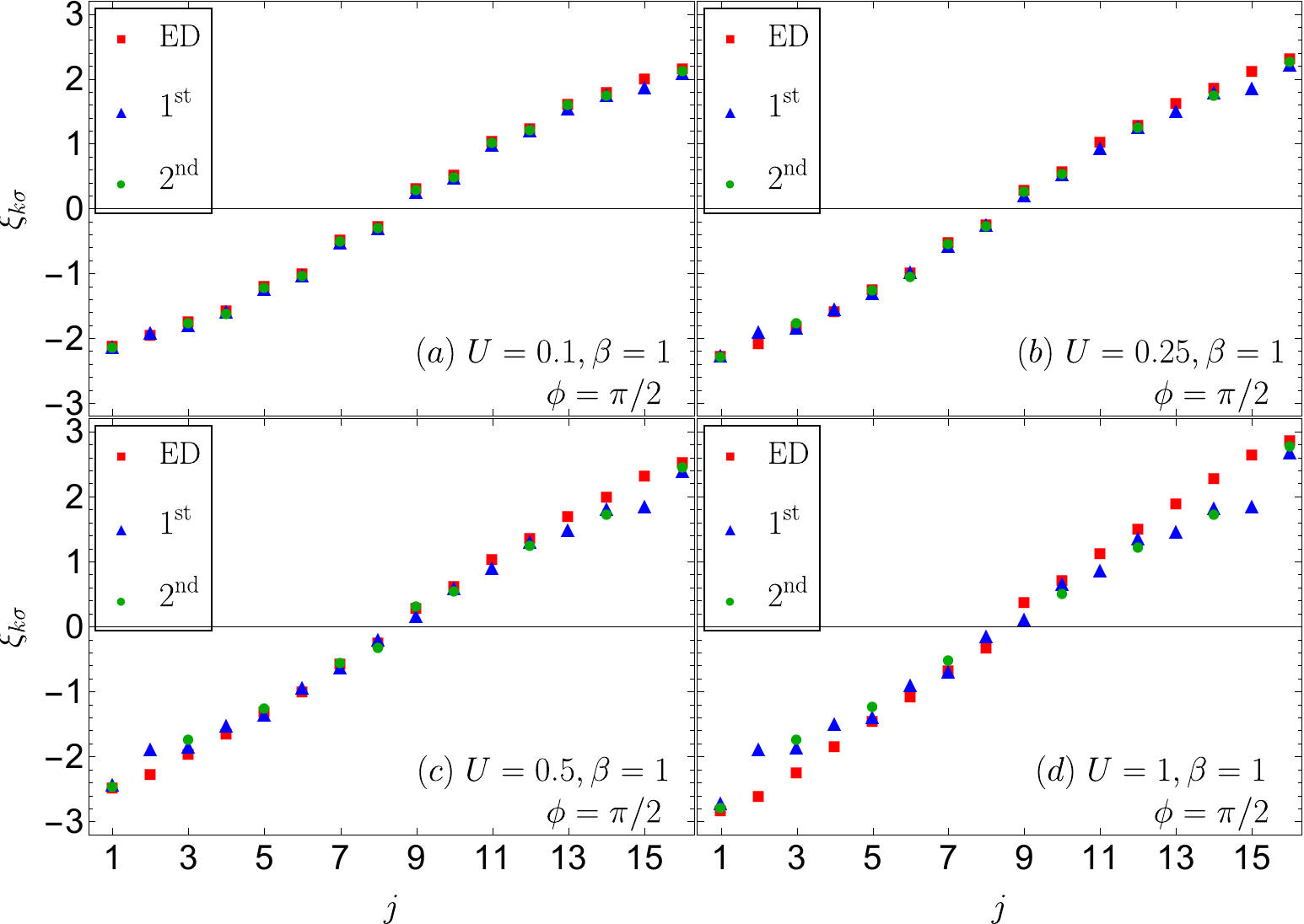}
    \caption{\footnotesize Comparison between the virtual energies $\xi_{k\sigma}$ of the 8-site Hubbard model with twisted BC's ($\phi=\pi/2$) found by ED(red squares), Eq.(\ref{eqn:theorem-1st-order-xi})(blue triangles), and Eq.(\ref{eqn:theorem-1st-order-xi})(green dots) at $\beta=1$ and different values of repulsive interaction $U>0$ marked on each plot. The twisted BC's lifted the degeneracies in the non-interacting energies, and hence, increased the agreement between ED and the perturbative approach, when compared to Fig.~\ref{fig:perturb_roots_periodic_repulsive}. On increasing $U>0$, more second-order solutions split into complex-conjugate pairs as seen for $U=1$, where every other solution is missing on the plot.}
\label{fig:perturb_roots_phi_1.57-repulsive}
\end{figure}
\begin{figure}[h!]
 \centering
    \includegraphics[width=\columnwidth]{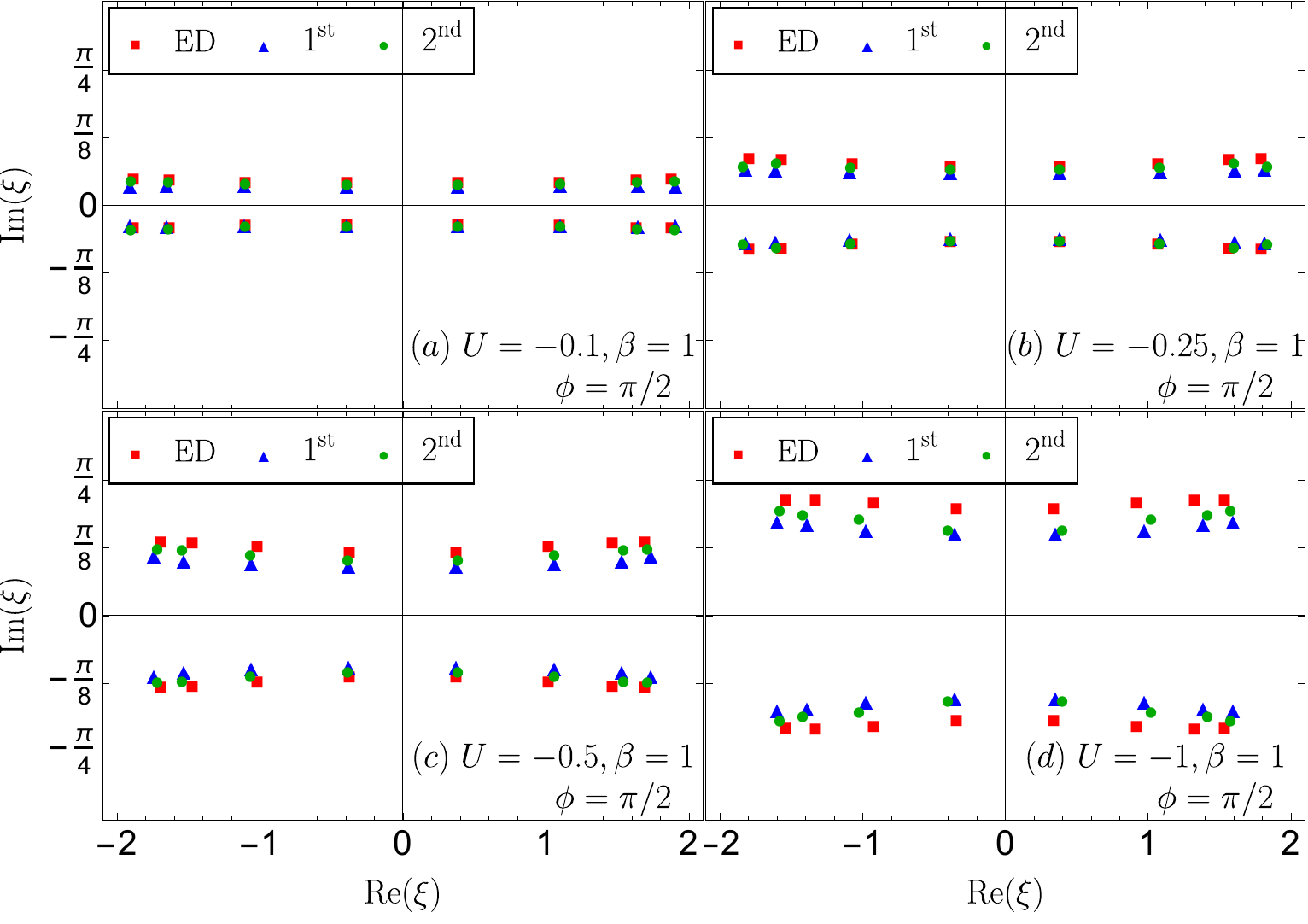}
    \caption{\footnotesize Comparison between the virtual energies $\xi_{k\sigma}$ of the 8-site Hubbard model with twisted BC's ($\phi=\pi/2$) at $\beta=1$ and different values of attractive interaction $U<0$ (markers and colors are the same as in Fig.~\ref{fig:perturb_roots_phi_1.57-repulsive}). In addition, the perturbative approach captures all the ED virtual energies as no degenerate energy levels are present. Compared to Fig.~\ref{fig:perturb_roots_periodic_attractive}, no displacement is needed, and better agreement with ED is observed. }
\label{fig:perturb_roots_phi_1.57-attractive}
\end{figure}

Comparisons between the virtual energies of the 8-site Hubbard model found by ED, first-order and second-order equations for $\phi=\pi/2$  for $U>0$ and $U<0$ are shown in Figs. \ref{fig:perturb_roots_phi_1.57-repulsive} and \ref{fig:perturb_roots_phi_1.57-attractive}, respectively. For most data points, the second-order virtual energies are closer to ED than the first-order virtual energies, and both worsen with increasing $|U|$. Since the energy levels are not degenerate, the perturbative approach is able to capture all the ED virtual energies. In the repulsive case, upon increasing $U$, we observe that more second-order solutions split into complex-conjugate pairs as seen for \ref{fig:perturb_roots_phi_1.57-repulsive}(b-d).

The results for different boundary conditions suggest a trade-off between degeneracy and the quality of the approximation for the second-order solutions. On one hand, the degenerate energies are more well-behaved as a function of $U$, where the number of solutions remains the same for all $U$ and the $U_\text{eff}$ is closer to the true $U$. On the other hand, the degenerate solutions fail to differentiate between the two exact diagonalization virtual energies corresponding to the same momentum label $k$, whereas the non-degenerate solutions capture this splitting better.  

\section{Charge fluctuations in a Hubbard ring coupled to an external charge probe}
\label{sec:charge-fluctuations}

We next consider a situation that might be amenable to future experimental study using scanning tunneling microscopy probes of the clusters of correlated atoms,   described as quantum corrals in \cite{eigler1990positioning, crommie1993imaging,crommie1993confinement,moon2009quantum}.

It is useful to introduce a simple decoherence-type model, parallel to central-spin models studied earlier \cite{Zurek1982Environment}. Consider a Hubbard ring of radius $R$ with $N_s$ atoms on the surface of a metal surface, which acts as a reservoir of charge, with an external charge probe, say an STM tip at the center. The total grand canonical Hamiltonian can be written as
\beq
H_{\text{tot}}= H_{\text{Hub}}- \mu \hat{N} + g \times \hat{n}_{tip} \; \hat{N}, \label{STM}
\eeq
where $\hat{N}$ is the total number operator for the Hubbard system, $\hat{n}_{tip}=0,1$ is the number of electrons on the external tip,  the last term represents the interaction of the Hubbard ring with the tip with $g=\frac{e^2}{R \varepsilon_0}$ and $R$ is the radius of the Hubbard ring. This setup allows the number of electrons on the Hubbard ring to fluctuate around a mean value determined by some gate voltage that translates to $\mu_0$, the real chemical potential, and the tip is then used to monitor these charge fluctuations.

We start the system at $t=0$ in a simple product state
\beq
|\Psi(0)\rangle= \left(\alpha |0\rangle+\beta |1\rangle\right)\otimes \sum_{N,\nu} C_{N, \nu} |N, \nu\rangle
\eeq
here $|0\rangle,|1\rangle$ are the states with external charge $0,1$ respectively, where  $|N,\nu\rangle$ and $E_{N,\nu}$ are the Hubbard eigenstates with $N$ electrons and $\nu$ represents the excitation label. The state at finite time can be found easily from $|\Psi(t)\rangle=e^{- i \frac{t}{\hbar} H_\text{tot}} |\Psi(0)\rangle\rangle$  as so that
\beq
|\Psi(t)\rangle=\sum_{N,\nu}e^{- i \frac{t}{\hbar} E_{N,\nu}}C_{N,\nu} \left(\alpha |0\rangle+\beta e^{- i g \frac{t}{\hbar} N } |1\rangle\right) \otimes |N,\nu\rangle \nn\\
\eeq
We form the reduced density matrix for the tip by tracing over the Hubbard states, assuming a thermal distribution of states:
\begin{eqnarray}
\hat{\rho}_{tip}&=&\frac{1}{Z(\mu_0)}\sum_{N,\nu}e^{- \beta(E_{N,\nu}-\mu_0 N} |\Psi(t)\rangle \langle \Psi(t)| \nn \\
&=& |\alpha|^2 |0\rangle \langle 0 |+ |\beta|^2 |1\rangle \langle 1| \nn \\
  & & +  (\alpha^* \beta) \, r(t) |1\rangle \langle 0|+(\alpha \beta^*) \, r^*(t) |0 \rangle \langle 1 |
\end{eqnarray}
where the decoherence factor
\beq
r(t, \mu_0)= \frac{1}{Z(\mu_0)} \tr e^{- \beta (H_{Hub} - \mu_0 \hat{N})} e^{- i \tau \hat{N}}
\label{eqn:r-of-tau}
\eeq
where
\beq
\tau=\frac{t}{\hbar} g.
\eeq
\begin{figure}[t]
 \centering
 \includegraphics[width=\columnwidth]{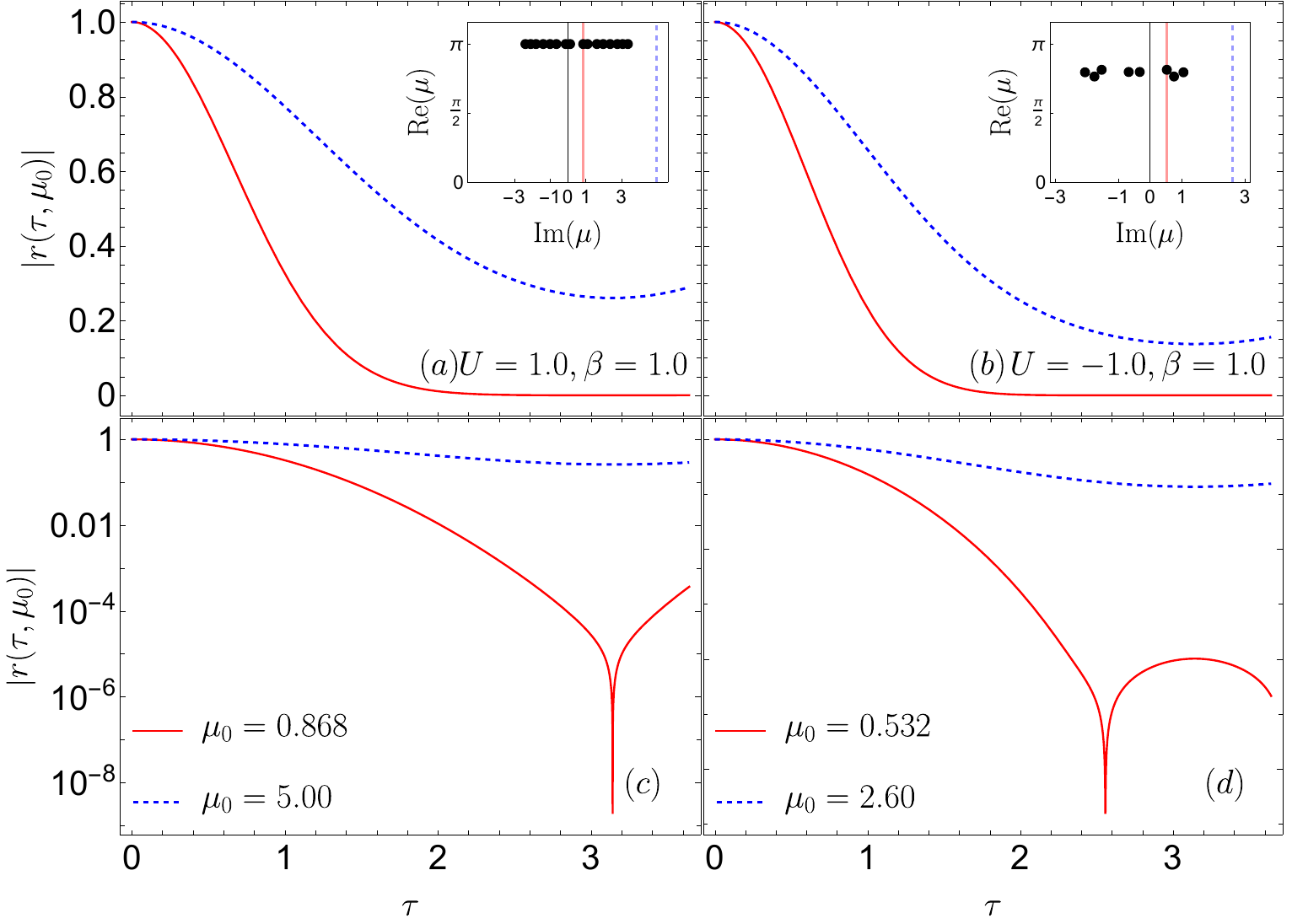}
 \caption{\footnotesize Correspondence between Yang-Lee zeros and times $\tau$ at which $|r(\tau,\mu_0)|$ vanishes from Eq.~(\ref{eqn:r-of-tau}). The results are for the 1-d Hubbard model with $N_s=8$, $\beta=1$ and $U=\pm 1$, indicated on the plot, for two different real chemical potentials $\mu_0$. Plots (a-b) show the regular scale of $|r(\tau,\mu_0)$ while (c-d) show the log scale of the same plots. The inset shows Yang-Lee zero in the complex $\mu$-plane. In (a), for example, the inset shows a complex zero at $\mu=0.868 + i \pi$ and the corresponding solid red curve of $|r(\tau,\mu_0)|$ vanishes at one time ($\tau =\pi$), indicating this root. However, for $\mu_0=5.00$, there are no zeros on the dashed blue line on the right panel, and the corresponding dashed blue curve of $|r(\tau,\mu_0)|$ does not vanish for any value of $\tau$. Similar behavior is observed for (b), where $U=-1$. In (c-d), we show the log scale of the plots in (a-b), respectively, to distinguish between the regions where the $|r(\tau,\mu_0)|$ and the point at which it vanishes.}
 \label{fig:r_object}
 \end{figure}
The object $r$ therefore contains the time-dependent charge fluctuations of the Hubbard model, and might in principle be obtained from measurements of charge fluctuations involving the off-diagonal elements of the reduced density matrix $\hat{\rho}_{tip}$.
We can calculate $r$ from the ED results. It is clear that $\beta \mu_0+ i \tau$ can be regarded as a complex chemical potential, and the zeros of $r$ in this complex variable correspond to the partition function zeros.

We use exact diagonalization of the 8-site Hubbard model with periodic BC's to plot $|r(\tau,\mu_0)|$ as a function of $\tau$ and for different values of $\mu_0$ and $U$ at $\beta=1$ in Fig.~\ref{fig:r_object}. The plot shows regular scale in  Fig.~\ref{fig:r_object}(a-b) and log scale in  Fig.~\ref{fig:r_object}(c-d) for the same values of $\mu_0$ and $U$, in addition to an inset of the complex $\mu$ plane for reference on where Yang-Lee zeros lie. We can see that for values of $\mu_0$ where there is a zero of the partition function, $|r(\tau,\mu_0)|$ vanishes when $\tau$ equals the imaginary part of the zero. The log scale in Fig.~\ref{fig:r_object}(c-d) is used to distinguish the point(s) at which the $|r(\tau,\mu_0)|$ vanishes from the neighborhood where $|r(\tau,\mu_0)|$ is very small. For situations where $|r(\tau,\mu_0)|$ is too small to pin down Yang-Lee zeros, it is beneficial to determine the regions of the complex $\mu$-plane where Yang-Lee zeros exist. 

\section{Conclusion}
\label{sec:conclusion}
In this work, we have extended the recently introduced perturbative technique of locating Yang-Lee zeros, or equivalently, the thermodynamic virtual energies, by finding the zeros of the inverse Green's function for a given momentum $k$ and spin  $\sigma$ to $O(U^2)$. The perturbative solutions up to $O(U)$ and $O(U^2)$ with various boundary conditions can be quantitatively compared by using an exact sum-rule \disp{Ueff,eqn:U_eff}, which is satisfied by the exact diagonalization results.

In Figure~\ref{fig:U-eff-periodic}, we see that for a given boundary condition, the solutions to $O(U^2)$ satisfy the sum-rules better than the solutions to $O(U)$. We also see that the twisted boundary conditions lead to better agreement between ED results and perturbative results due to lifting the degeneracy in the momentum label, as seen in Figures~\ref{fig:perturb_roots_periodic_repulsive},\ref{fig:perturb_roots_periodic_attractive},\ref{fig:perturb_roots_phi_1.57-repulsive}, and \ref{fig:perturb_roots_phi_1.57-attractive}.

We also suggest that it should be possible to experimentally study these virtual energies using time-dependent charge fluctuations of the Hubbard model. The suggested setup involves STM studies of clusters of correlated atoms in quantum corrals \cite{eigler1990positioning, crommie1993imaging,crommie1993confinement,moon2009quantum}, which seem to be available.  Measurement of the tunneling current  fluctuations, in a the tip of a scanning tunneling microscope could lead to experimental insights into this topic.

\clearpage
 \bibliography{References}

\begin{thebibliography}{10}
\bibitem{footnote}
When labeling the real solutions with spin projection quantum number $\sigma$, we use a convention where the solution with lower (higher) real part of $\xi$ is regarded as arising from spin up (down). Similarly, when both solutions are complex conjugate, the solution with a positive(negative) imaginary part of $\xi$ is regarded as arising from spin up (down).
\end{thebibliography}
 \bibliographystyle{ieeetr}


\end{document}